\documentstyle[12pt,aas2pp4,flushrt,epsf]{article}    

\begin{document}
\title{Violent Relaxation of Indistinguishable Objects and\\ 
Neutrino Hot Dark Matter in Clusters of Galaxies}
\author{Kull A., Treumann R. A., and B\"ohringer H.}
\affil{Max-Planck-Institut f\"ur Extraterrestrische Physik\\
        D-85740 Garching, Germany}

\begin{abstract}
The statistical mechanical investigation of violent relaxation (Lynden-Bell
1967) is extended to indistinguishable objects. It is found that, 
coincidentally, the equilibrium distribution is the same as that obtained for 
classical objects. For massive neutrinos, the Tremaine \& Gunn (1979) phase 
space bound is revisited and reinterpretated as the limit indicating the onset 
of degeneracy related to the coarse-grained phase space distribution. In the 
context of one of the currently most popular cosmological models, the Cold and 
Hot Dark Matter (CHDM) model (Primack et al. 1995), the onset of degeneracy 
may be of importance in the core region of clusters of galaxies. Degeneracy 
allows the neutrino HDM density to exceed the limit imposed by the Tremaine 
\& Gunn (1979) bound while accounting for the phase space bound.
\end{abstract}


\section{Introduction}
Self-gravitating astrophysical systems having initial mass and velocity 
distributions far from equilibrium are believed to evolve towards equilibrium 
by violent relaxation, a collective, collisionless process mediated by the
fluctuating gravitational field. Compared to the time scale of the two-body 
relaxation process, the time scale of violent relaxation is short.
The analytical investigation of violent relaxation in terms of statistical 
mechanics was first introduced by Lynden-Bell (1967). Formally, 
the equilibrium distribution resembles the Fermi-Dirac distribution. 
For stellar and galactic systems it tends towards the 
Maxwell-Boltzmann distribution (Lynden-Bell 1967, Shu 1978).

In this Letter we investigate violent relaxation of indistinguishable objects. 
We show that, despite of the difference between the violent relaxation statistics of 
indistinguishable and distinguishable objects, coincidentally, the equilibrium
distributions of both cases are of the same form as expected by Shu (1987). 
When applied to massive neutrinos it turns out that the degeneracy implicit in the 
equilibrium distribution plays an important role leading to a reinterpretation of the 
Tremaine \& Gunn (1979) phase space bound as a limit indicating the onset of 
degeneracy. In the context of one of the currently 
most popular cosmological models based on initial conditions compatible with 
inflation theory ($\Omega=1$), the Cold and Hot Dark Matter (CHDM) model 
(Primack et al. 1995), the onset of degeneracy allows the neutrino HDM
density to exceed the limit imposed by the Tremaine \& Gunn (1979) bound
while accounting for the phase space bound.

\section{Violent Relaxation Statistics of Indistinguishable Objects}
Following Lynden-Bell (1967) we attempt to describe violent relaxation
by the use of a maximum entropy principle. The one-particle phase space 
$\mu$ of the objects undergoing violent relaxation is divided into a 
large number of microcells the dimension of which is chosen so as to yield the 
mass density in the units mass per spatial and velocity volume, $\Delta^3x$ 
and $\Delta^3v$, respectively. The collisionless
nature of violent relaxation then imposes an additional constraint on the
time evolution of the $\mu$-space. Initially not overlapping phase elements
do never overlap. Therefore, a microscopic exclusion principle for phase 
elements in $\mu$-space is established.
The volume of the microcells is chosen in such a way that the
microcells are occupied by at most one object each and that
the probability of finding occupied adjacent microcells 
representing two-body encounters is small (Saslaw 1985). With $m$ 
the particle mass, the mass per microcell is either zero or $m$. 
Let $\eta$ denote the corresponding mass density of an occupied microcell.
The initial phase space distribution before violent relaxation
may then be described by the set of occupied microcells.
 
At the macroscopic level, the microcells (or phase 
elements) are grouped into macrocells containing a large number $\nu$ of 
microcells. Let $N$ be the total number of occupied microcells corresponding 
to $N$ objects. Suppose that there is a microstate in which there are $n_a$ 
occupied microcells in the $a$th macrocell. Due to the collisionless 
interaction of the violent relaxation process, the occupation state of 
each microcell does not change, i.e. there is no cohabitation. The 
indistinguishable nature of the objects under consideration leads to a 
total number
\begin{equation}
w_a={\nu!\over{n_a!(\nu-n_a)!}}
\end{equation}
of ways of assigning the $n_a$ occupied microcells to the $\nu$ microcells of
the $a$th macrocell as already realized by Lynden-Bell (1967) but not worked out 
further. Since the elements are indistinguishable, there is just
one way of splitting the total of $N$ elements into groups $n_a$, i.e. there 
are no permutations. Thus, for indistinguishable elements, the total number of 
microstates is
\begin{equation}
W_i=\prod_a {\nu!\over{n_a!(\nu-n_a)!}}
\end{equation}
where, for distinguishable elements, the corresponding number is 
(Lynden-Bell 1967)
\begin{equation}
W_d={N!\over{\prod_a n_a!}} \prod_a {\nu!\over{(\nu-n_a)!}} \,.
\end{equation}

The most probable state is found by the standard procedure of maximizing
log $W$ subject to the constraints of constant total energy and constant mass. 
Since the expressions $W_i$ and $W_d$ differ only by a factor $N!$, the
maximization procedure performed in the continuum representation yields, 
coincidentally, for both expressions log $W_i$ and log $W_d$ the same 
coarse-grained phase space distribution
\begin{equation}
\bar{f}(v,x)={\eta \over{1+\mbox{exp}[\beta(\epsilon - \mu)]}} \,.
\label{lbdist}
\end{equation}
Here, $\epsilon=v^2/2+\Phi(x)$, with $\Phi(x)$ the gravitational 
potential normalized as $\Phi(x \rightarrow \infty)=0$. 
The Lagrange parameters $\beta$ and $\mu$ are chosen according to 
the macroscopic constraints.
Thus, the most probable coarse-grained phase space distribution of 
indistinguishable elements is the same as for distinguishable elements.
 
Shu (1978) reexamined the statistical mechanical discussion of violent relaxation 
in terms of particles and obtained an expression equivalent to (\ref{lbdist})
determining the occupation number $n_a$  
\begin{equation}
n_a={\nu \over{1+\mbox{exp}[\beta_p m (\epsilon_a - \mu_p)]}}
\label{shudist}
\end{equation}
of the $a$th macrocell where $\epsilon_a=v_a^2/2+\Phi(x_a)$, and the Lagrange 
parameters, $\beta_p$ and $\mu_p$, now refer to the particle description. For 
both the cases (\ref{lbdist}) and (\ref{shudist}), with $\bar{f} \ll \eta$ and 
$\nu_a \ll \nu$, respectively, the phase space distribution tends towards 
the Maxwell-Boltzmann distribution.
Despite of these agreements, the discussions by Shu (1978) and Lynden-Bell (1967) 
differ what concerns degeneracy effects potentially present in (\ref{lbdist}) 
and (\ref{shudist}). While Lynden-Bell (1967) considers degeneracy to be eventually 
important in central regions of galaxies, Shu (1978) concludes degeneracy to 
indicate the onset of two-body encounters. However, if two-body encounters are 
present, there is no microscopic exclusion principle any more. Thus the relevant  
phase space distribution is of Maxwell-Boltzmann type. In any case, 
for massive neutrinos, the discussion turns out to be irrelevant.

\section{The Case of Massive Neutrinos}
If the neutrino has a non-zero rest mass as predicted by some extensions of 
the Standard Model (Gelmini \& Roulet, 1995), this may have important consequences 
for both the solar neutrino and the cosmological dark matter problem. Massive 
neutrinos with a rest mass of a few eV have been shown to account remarkably 
well for both problems (Primack et al. 1995). The discussion in this section 
is restricted to one neutrino flavor with $g_\nu=1$.

We now apply the result of the foregoing section to relict neutrinos of 
non-zero rest mass trapped in a gravitational potential, e.g. related to a cluster 
of galaxies. The initial phase space density of the unperturbed relict 
neutrino background is a relativistic Fermi distribution 
\begin{equation}
f_\nu(p)={m_\nu^4 \over{(2 \pi \hbar)^3}}
{1 \over {e^{(p c/{kT_\nu})}+1}}
\label{initdist}
\end{equation}
with $T_\nu=1.95$ K (for $z=0$ and in the units mass per spatial and 
velocity volume $d^3x$ and $d^3v$, respectively). For $p=0$, the 
maximum value $f_\nu(0)=(1/2)\,m_\nu^4/(2 \pi \hbar)^3$ is obtained. 

If the neutrinos are massive, the phase space distribution $f$ of the
unperturbed relict neutrino background is changed due to gravitational
interactions. Violent relaxation considered here leads to a convolution 
of the phase space structure. Regions of initially different 
phase space densities end up entwined together. However, the 
coarse-grained phase space density $\bar{f}$ cannot exceed the maximum 
of the initial fine-grained phase space density $f_\nu$ since the 
collisionless nature of violent relaxation implies conservation of the 
fine-grained phase space density $f$. 
 
Since neutrinos are elementary particles and their unperturbed, initial 
phase space distribution is the relativistic Fermi distribution 
(\ref{initdist}), the volume of the microcells is chosen to be the 
elementary phase space volume $m_\nu^3/(2 \pi \hbar)^3$. This 
choice guarantees the microcells to be occupied by at most one object 
each. The corresponding phase space mass density of an occupied microcell 
is $\eta=m_\nu^4/(2 \pi \hbar)^3$. Note, that the initial condition assumed 
in the foregoing section is perfectly fulfilled by the phase space 
distribution of the relict neutrinos.

Suppose now the relict massive neutrinos to be subject to violent relaxation.
According to the foregoing section, their coarse-grained phase space distribution 
$\bar{f}$ tends towards the Fermi type distribution (\ref{lbdist}). 
First we note, that, in general, the relation max$\{\bar{f}\}\le$ max$\{f\}$ must hold.
With the natural assumption $\beta > 0$, equation (\ref{lbdist}) implies
\begin{equation}
\left(\Phi - \mu\right) \ge 0 \,.
\label{inequ0}
\end{equation}  
where 
for a given $\beta$ equality corresponds to the highest possible phase space 
density. If no degeneracy effects are present, 
then $\bar{f} \ll \eta$ and equation (\ref{lbdist}) requires
\begin{equation}
\beta\left(\Phi - \mu\right) \gg 0 \,.
\label{inequ1}
\end{equation}
In terms of the mass density
\begin{equation}
\rho_\nu= 4\pi\eta\int\limits_{0}^{\infty}v^2
	e^{-\beta\left[(v^2/2+\Phi)-\mu\right]}dv
\end{equation}
and the specified value of $\eta=m_\nu^4/(2 \pi \hbar)^3$, 
inequality (\ref{inequ1}) leads to 
\begin{equation}
\rho_\nu \ll {m_\nu^4 \sigma^3\over{\sqrt{8 \pi^3}\hbar^3}}
\label{inequ2}
\end{equation}
where $\sigma = \beta^{-1/2}$ is the one-dimensional velocity dispersion. 
When solved for $m_\nu$ we get
\begin{equation}
m_\nu \gg \left( {\rho_\nu \sqrt{8 \pi^3}\hbar^3 \over{\sigma^3}} \right)^{1/4} 
\doteq m_{d}
\label{inequ3}
\end{equation}
This expression is interpreted as a limit on the type of the
coarse-grained phase space density $\bar{f}$. For given velocity 
dispersion $\sigma$, neutrino density $\rho_\nu$ and a neutrino mass $m_\nu$ 
well above $m_{d}$, $\bar{f}$ is the Maxwell-Boltzmann distribution. On the 
other hand, if $m_\nu$ is of order or even below $m_{d}$, $\bar{f}$ differs
from the Maxwell-Boltzmann distribution. In this case, degeneracy effects
become important and the correct coarse-grained phase space density $\bar{f}$
is the Fermi-type distribution (\ref{lbdist}). We note, that the possible
onset of two-body encounters preventing Fermi-like degeneracy in the case
of stars or galaxies (Shu 1978) does not apply to neutrinos. The time scale
of gravitational two-body encounters for neutrinos is extremely long. Moreover,
even when strong two-body encounters would occur, the Fermionic nature of
the neutrinos prevents single microcells to become cohabitated. Thus,
for neutrinos, a microscopic exclusion principle applies in any case.

Inequality (\ref{inequ3}) is formally equivalent to the well known Tremaine \& Gunn 
(1979) bound on the rest mass of neutrinos which are gravitationally bound in 
galactic halos or galaxy clusters. However, the Tremaine \& Gunn (1979) limit 
is derived from the assumption that $\bar{f}$ is the Maxwell-Boltzmann distribution
and the condition that the maximum of the coarse-grained phase space density 
$\bar{f}$ cannot exceed the maximum of the initial fine-grained phase space density 
$f$ given by (\ref{initdist}). Therefore, the Tremaine \& Gunn (1979) limit 
implicitly relies on the assumption of $\bar{f}$ to be of Maxwell-Boltzmann type.
On the contrary, (\ref{inequ3}) refers to the type of
the phase space distribution. Therefore we stress,
that, from the point of view of (\ref{inequ3}), the limit implied should not 
be interpreted as limit on the neutrino mass. Instead, (\ref{inequ3}) states 
that the phase space distribution  for $m_\nu$ near the Tremaine \& Gunn (1979) 
bound differs significantly from the Maxwell-Boltzmann distribution. 

In terms of the gravitational potential $\Phi$ and the neutrino mass
$m_\nu$ it is possible to formulate several limits on the neutrino rest 
mass $m_\nu$ based on the phase space density $\bar{f}$ defined by 
(\ref{lbdist}). We intend here to derive the most robust one. 
According to (\ref{inequ0}) the mass density of the totally degenerate, 
densest state is
\begin{equation}
\rho_\nu= 4\pi\eta/2\int\limits_{0}^{v_l} v^2 dv
\label{degstat}
\end{equation}
where $v_l$ is the limiting velocity below which the maximum number of 
microcells is occupied ($\bar{f}=\eta/2$) while above  $v_l$ all microcells 
are empty ($\bar{f}=0$). Assuming only bound states to be relevant, i.e. 
$\epsilon = v^2/2 + \Phi \le 0$, the most robust limit on the neutrino mass
is found for $v_l = \sqrt{|2 \Phi|}$ yielding
\begin{equation}
m_\nu \ge \left({12 \rho_\nu \pi^2 \hbar^3 \over{|2 \Phi|^{3/2}}}\right)^{1/4}.
\label{limit1}
\end{equation}
This limit is robust in the sense that it holds for every coarse
grained phase space distribution $\bar{f}$ representing a bound state including
those eventually not described by (\ref{lbdist}), as for example 
anisotropic phase space distributions suggested by Madsen (1991) or
Ralstone \& Smith (1991).
In order to compare these limits to the Tremaine \& Gunn (1979) bound, the relation 
$\sigma^2 \approx \Phi/3$ holding for a Maxwellian velocity distribution 
related to a self-gravitating isothermal sphere is adopted. 
The Tremaine \& Gunn (1979) bound on the neutrino rest mass then becomes 
\begin{equation}
m_\nu \ge \rho_\nu^{1/4} \left({6 \pi \hbar^2 \over{|\Phi|}}\right)^{3/8}
\label{limit2}
\end{equation}
which is about a factor $\approx 1.18$ higher than (\ref{limit1}). In terms of
the neutrino density this leads to an augmentation of the density limit of about 
a factor of $\approx 1.95$. As a consequence, the onset of degeneracy allows the 
neutrino HDM density to exceed the limit imposed by the Tremaine \& Gunn (1979) 
bound by a factor of $\approx 1.95$ while accounting for the phase space bound. 

In comparing the two limits, one should take into account that the Maxwellian
velocity distribution related to (\ref{limit2}) has a tail of unbounded 
neutrinos. For the probably more physical truncated Maxwellian distribution 
consisting of bounded neutrinos only, the difference between the limits would
become bigger. On the other hand, the totally degenerate situation on which the 
limit (\ref{limit1}) is based is unlikely to be realized by violent relaxation. 

\section{Discussion and Conclusion}
This section considers violent relaxation of massive neutrinos in
clusters of galaxies in the context of the currently popular Cold and Hot Dark 
Matter (CHDM) model. Extensive numerical simulations show these models to agree 
well with observations on cosmological scales (e.g. Klypin et al. 1993,
Noltenius et al. 1994, Klypin \& Rhee 1994). In what follows we refer to the two most 
promising neutrino mass schemes due to Primack et al. (1995). The first scheme divides 
the neutrino HDM in two neutrino species each with a mass of 2.4 eV while the second 
splits the HDM in three 1.6 eV neutrino species. The median value of the
initial phase space distribution $f_\nu$ of each family of the relict neutrinos
is only $0.06$ (Madsen \& Epstein 1984).  As an approximation we therefore again
take $\eta=m_\nu^4/(2 \pi \hbar)^3$ where $m_\nu$ is 2.4 eV or 1.6 eV,
respectively, i.e. for our purposes the phase space elements occupied by two or 
three neutrinos are neglected.

The left panel of figure \ref{fig1} presents the contour plot of density limits 
derived from (\ref{limit1}) and (\ref{inequ2}) for the 1.6 eV neutrino mass scheme. 
The right panel shows the density limits for the 2.4 eV mass scheme. The limits are 
plotted as upper limits to the total matter densities $\rho$ consisting of HDM, CDM 
and baryonic matter. They are functions of the gravitational potential depth, 
indicated by the velocity dispersion of the corresponding self-gravitating isothermal 
sphere $\sigma \approx \sqrt{\Phi/3}$, and the ratio of the neutrino 
HDM density $\rho_\nu$ to the total matter density $\rho$. The solid contours represent 
the most robust limit for bound states given by (\ref{limit1}). Along these contours 
of specified total density $\rho$, the neutrino HDM is found in the totally degenerate, 
densest state. Augmentation of $\rho$ is only possible if it is accompanied by a 
change of the ratio $\rho_\nu/\rho$ or $\sigma$. On the other hand, the more $\rho_\nu/\rho$ 
is diminished while leaving $\rho$ and $\sigma$ unchanged, the more the coarse-grained
phase space $\bar{f}$ of the neutrino HDM becomes non-degenerate. The dashed contours
are the density limit obtained from (\ref{inequ2}). Well below this limit, $\bar{f}$
becomes a Maxwellian. In addition, this density limit by (\ref{limit2}) corresponds
to the Tremaine \& Gunn (1979) bound. 

The parameter range of figure \ref{fig1} includes typical values of $\sigma$ and 
$\rho_\nu/\rho$ expected for clusters of galaxies. The suppression of the HDM neutrino 
density $\rho_\nu$ compared to the CDM and baryonic matter density due to its lower
density contrasts (e.g. Klypin et al. 1993, Noltenius et al. 1994)
is reflected by values of $\rho_\nu/\rho$ below the cosmological value $\Omega_\nu=0.2$ 
(Primack et al. 1995). For the 1.6 eV neutrino mass scheme (left panel of 
figure \ref{fig1}), one observes a broad parameter region yielding upper total 
density limits comparable to typical core densities of cluster of galaxies. 
Adopting the 2.4 eV neutrino mass scheme (right panel of figure \ref{fig1}) shrinks 
the relevant parameter region. However, since the coarse-grained phase space 
distribution $\bar{f}$ of the neutrino HDM is expected to be a Maxwellian only well 
below the limit indicated by the dashed contours, it is found that for both CHDM models 
$\bar{f}$ may differ significantly form the Maxwell-type distribution. Accordingly, 
$\bar{f}$ should be represented by the Fermi-type distribution (\ref{lbdist}). Due to 
(\ref{inequ0}) it accounts for the phase space bound while it allows the neutrino density 
to exceed the density limit imposed by the Tremaine \& Gunn (1979) bound 
(indicated by the dashed contour lines, too) by a factor of $\approx 1.95$.

Violent relaxation eventually fades before the final state (\ref{lbdist}) 
is attained. Thus, (\ref{lbdist}) is unlikely to be reached throughout the whole
cluster, but it is reasonable to hold in the central region where violent 
relaxation occurs most violently. In order to determine the global properties 
of self-gravitating spheres based on coarse grained phase space distributions 
(\ref{lbdist}), a careful investigation is needed. In particular, the onset of
degeneracy eventually related to core formation and not yet observed in simulations 
(Bryan et al. 1994) appears interesting.
  
In summary, the statistical mechanical investigation of violent relaxation 
has been extended to indistinguishable objects. It is found that the 
equilibrium distribution is the same as that obtained for distinguishable 
objects. However, in contrast to stellar and galactic systems, the onset of 
degeneracy is not prevented by two-body encounters in the case of massive 
neutrinos. Thus, the coarse-grained phase space density $\bar{f}$ is of 
Fermi-type and, to some extent, degeneracy may be present. For currently  
popular CHDM models, the onset of degeneracy is shown to be relevant in 
the core region of clusters of galaxies.

The authors thank Anatoly Klypin for valuable criticism and discussion.

\onecolumn
\newpage

\figcaption{Contour plot of upper limits to the total density $\rho$ (in kg m$^{-3}$) for
the 1.6 eV (left panel) and the 2.4 eV (right panel) neutrino mass scheme.
The limits are plotted as functions of the gravitational potential represented by 
the one dimensional velocity dispersion $\sigma$ according to ${\sigma^2 \approx \Phi/3}$ 
and the ratio $\rho_\nu/\rho$. The solid contour lines represent the
most robust density limit for bound states. Dashed contour lines indicate the density limit related
to the type of the coarse-grained phase space distribution $\bar{f}$ of the neutrino HDM. Well
below this limit $\bar{f}$ is a Maxwellian.
\label{fig1}}

\begin{figure}
\plotone{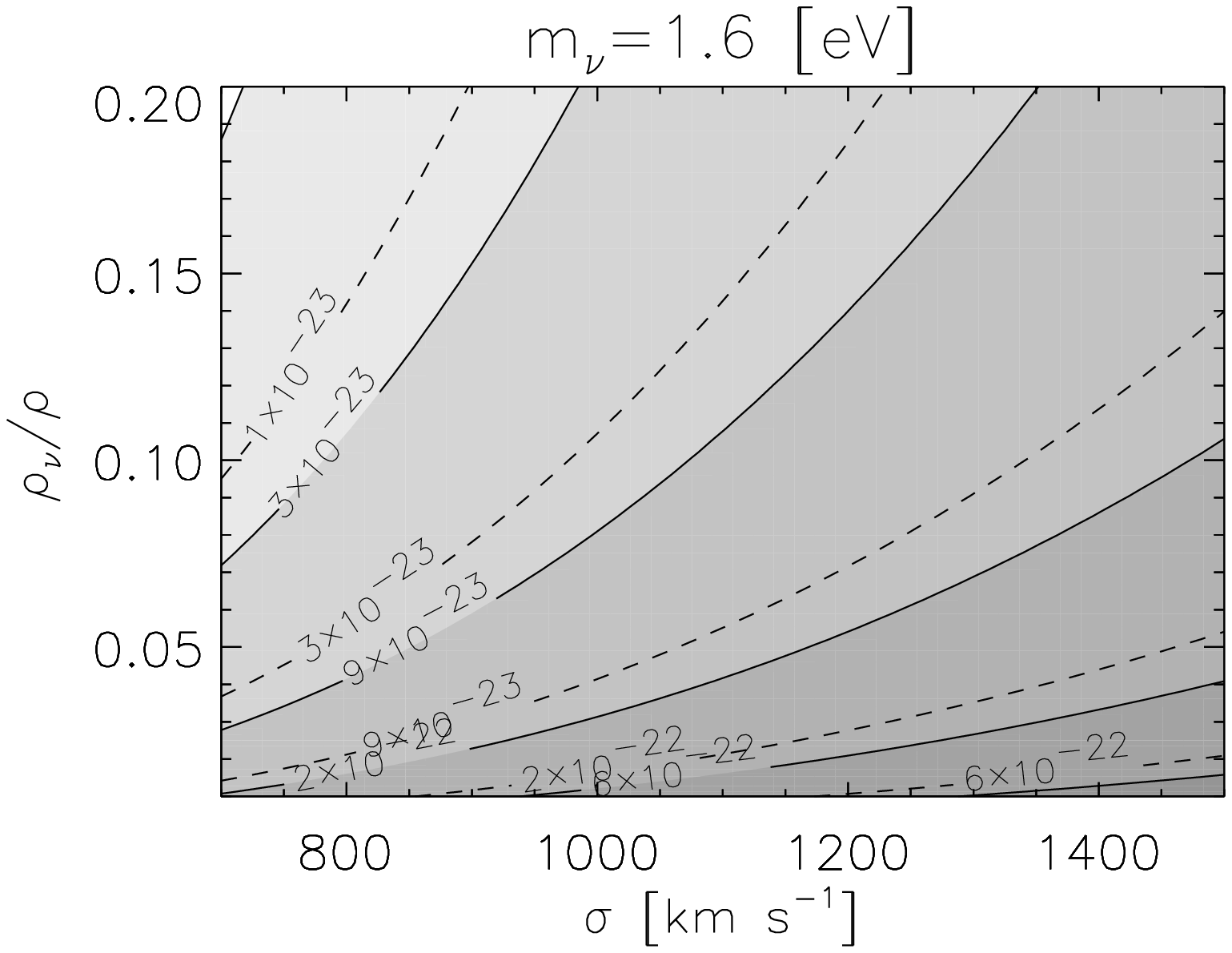}

Fig. 1.- (left panel)
\end{figure}
\begin{figure}
\plotone{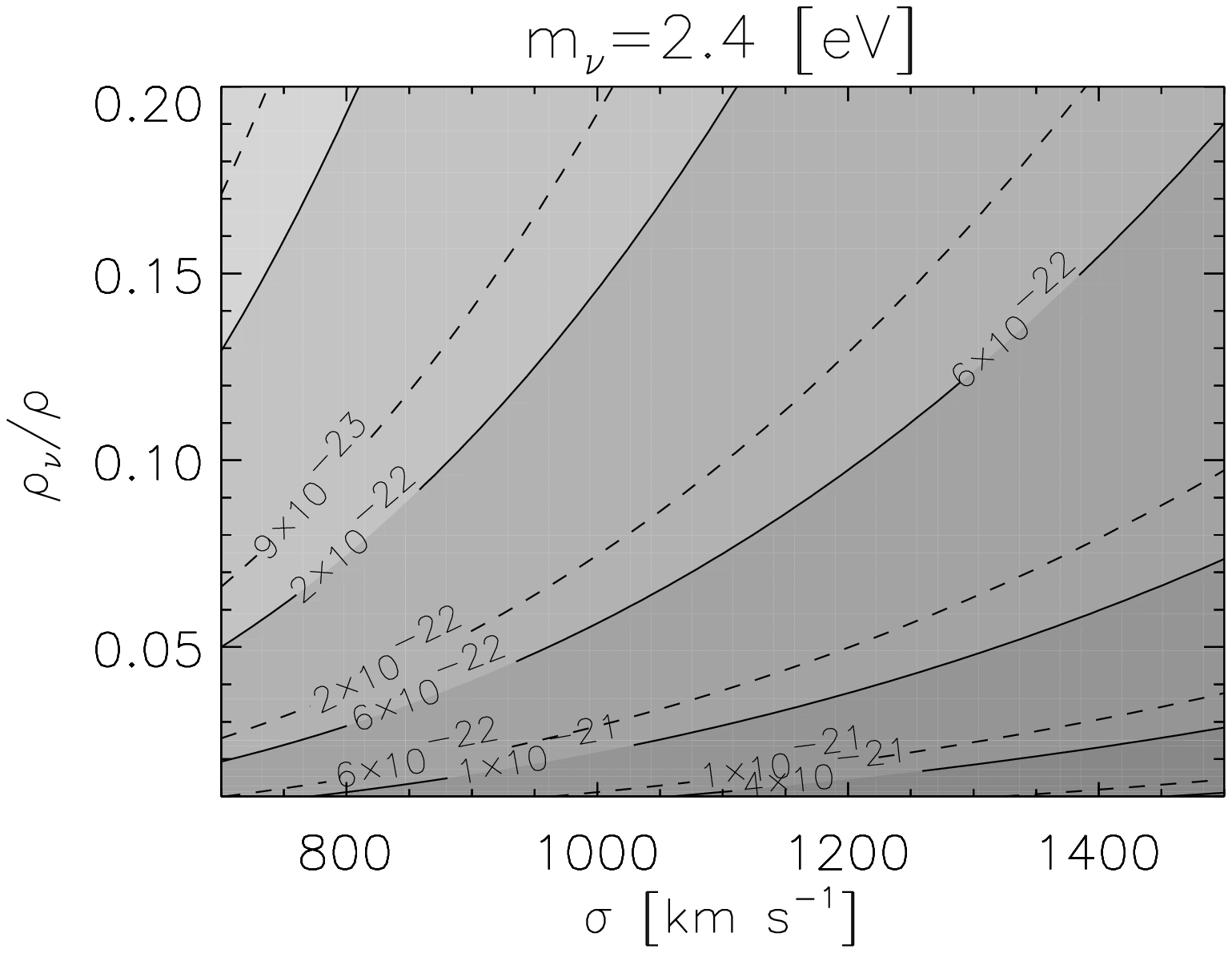}

Fig. 1.- (right panel)
\end{figure}

\end{document}